\documentclass[12pt]{iopart}

\usepackage{graphicx}
\usepackage{braket,iopams}

\begin{document}

\title[Strong coupling of two interacting excitons]{Strong coupling of two interacting excitons confined in a nanocavity-quantum-dot system}

\author{Paulo C. C\'ardenas$^1$, Nicol\'as Quesada$^2$, Herbert Vinck-Posada$^3$ and Boris A. Rodr\'iguez$^1$} 

\address{$^1$ Instituto de F\'isica, Universidad de Antioquia, Medell\'in AA 1226, Colombia \\
$^2$McLennan Physical Laboratories, University of Toronto, 60 St. George Street, Toronto, Ontario, Canada M5S 1A7\\
$^3$ Departamento de F\'isica, Universidad Nacional de Colombia, Ciudad Universitaria, Bogot\'a, Colombia
}

\ead{pcardenas@fisica.udea.edu.co}

\begin{abstract}
We present a study of the strong coupling between radiation and matter,
considering a system of two quantum dots, which are in mutual interaction and
interacting with a single mode of light confined in a semiconductor
nanocavity. We take into account dissipative mechanisms such
as the escape of the cavity photons, decay of the quantum dot excitons by
spontaneous emission, and independent exciton pumping. It is shown that the
mutual interaction between the dots can be measured off-resonance, only if the
strong coupling condition is reached. Using the quantum regression theorem, a reasonable definition of the dynamical coupling regimes is introduced in terms of the complex Rabi frequency. Finally, the emission spectrum for relevant conditions is presented and compared with the above definition, demonstrating that the interaction between the excitons does not affect the Strong Coupling. 
\end{abstract}


\submitto{\JPCM}

\maketitle

\section{Introduction}
\label{S1}
Cavity Quantum Electrodynamics (CQED) has provided an appropriate framework to 
understand the interaction between light and matter in a full quantum level.
One of the most relevant achievements of CQED is the coherent and
reversible exchange of energy between an active medium and the cavity mode.
This behavior is known as the Strong Coupling (SC) regime \cite{Gerry}. 
In high finesse QED cavities SC was achieved with Rydberg atoms 
several years ago \cite{Kimble,Haroche}. It was also realized in the last
decade in semiconductor systems, where a possible  
physical system consists of a quantum dot (QD) embedded in a
semiconductor micro(nano)-cavity \cite{Reith04,Yoshi,Bloch,Fin0}, in which the QD
discrete level structure resembles the atomic CQED physics.
The possibility of achieving SC in semiconductor systems, allows to
consider possible applications such as single photon sources \cite{Gerard,Yamamoto2}, 
coherence and entanglement control \cite{Nico}, quantum computation and quantum information
processing \cite{Yao,Imamo,Jeremy}, Bose-Einstein condensation of
polaritons \cite{Yamamoto1,Littlewood} or polariton lasing \cite{Bloch1,Vera,Nico2}.\\
Weak (WC) or strong (SC) coupling regimes for a QD microcavity system can be determined
in a micro-photoluminescence ($\mu$PL) experiment, in which the emission
spectrum of the micro(nano)-cavity is associated to the dynamical regimes of the
system \cite{Nico,Elena08}. In these experiments, two peaks are clearly
identified far off resonance. One of them is associated to the excitonic transition whereas the
second one is related to the cavity photons. Furthermore, near resonance the state
structure becomes more complex and the simple exciton-photon picture (WC) may change to the so-called 
polaritonic states. The splitting between the two peaks 
depends on the dissipation rates (decay and pumping), the coupling constants between the subsystems, and 
the detuning between the exciton and photon energies. Data obtained experimentally from
the $\mu$PL spectrum can be studied as a function of the detuning parameter
$\Delta$. In the resonance condition ($\Delta = 0$) if both peaks cross, then the system is in weak coupling, otherwise the system is in the SC regime \cite{Reith04,Yamamoto2}.\\
A model that is able to reproduce the above mentioned experimental
facts is presented in \cite{Elena08}. In this work, the $\mu$PL
spectrum of a QD micro(nano)-cavity system is modeled considering photonic 
and excitonic incoherent pumpings, and decay processes. In addition, different 
coupling regime conditions in the linear (low power excitation density) and non-linear 
(high power excitation density) were introduced by Tejedor and coworkers in \cite{Tej09,Tej09b}, recently a similar model extended for $N$ independent QDs coupled to a single common cavity mode (Dicke model) was presented in  \cite{DelValle2}. Besides these results, in \cite{Reitzenstein,Fin10} the $\mu$PL spectrum associated to the 
coupling of two semiconductor QDs to a single nanocavity mode was presented.\\ 
When more than one QD is considered they not only interact with the light mode, but also interact among themselves. The physics of this interaction could be used in quantum logic devices and quantum computation applications \cite{Imamo2}. 
One of the mechanisms in which two excitons can interact is through a resonant energy transfer, 
the so-called F\"orster interaction \cite{Lovett}. It has has been characterized 
in the experiments performed in \cite{Kag96,Klim02}. The aim of this paper is to show how 
the coupling regimes change due to the mutual interaction between the QDs, considering 
dissipative effects. Our main finding is that the mutual interaction between the QDs can be associated to 
a SC condition out of resonance. Nonetheless, we show that the WC and SC regimes do not change 
significantly, as a function of the mutual interaction strength between the QDs. This result opens the possibility of designing solid state all-optical quantum networks by deterministically growing QDs in nanocavities.\\
This paper is organized as follows: in section 2, we present the Hamiltonian and dissipative dynamics 
of the system. In section 3, we introduce the quantum regression theorem (QRT), which allows to 
compute two-time correlation functions and the emission spectrum of the system. 
Once the theoretical tools are introduced, in section 4 we go back to the Hamiltonian of the system, and obtain the energies and the polaritonic states (dressed states). Then, we show the contributions of each sub-system to the polaritonic states. Next, by including the dissipative effects 
in the dynamics of the system and using the QRT matrix, we introduce the complex (half) Rabi frequency 
which serves as a criterion for distinguishing the different coupling regimes. Finally to support our results we present the emission spectrum of the system. The discussion and conclusions of this work are presented in section 5.

\section{The System}
\label{S2}

The system considered here, is composed of two interacting and spatially separated QDs 
coupled to a single nanocavity mode of a high Q photonic crystal. We model the QDs as 
two level systems, and use a Tavis-Cummings like model. Furthermore, we include 
a F\"orster type interaction between the QDs. This allow us to write the following 
Hamiltonian (in units in which $\hbar = 1$):

\begin{equation}\label{eq.1}
\hat{H}_{S} = \omega_0 \hat{a}^{\dagger}
\hat{a}+ \sum_{i=1}^{2} \left\lbrace \omega_{X}
\left( \hat{\sigma}_i^{\dagger}\hat{\sigma}_i \right) + g_{i} \left(\hat{\sigma}^{\dagger}_{i} \hat{a} +
\hat{\sigma}_{i}\hat{a}^{\dagger} \right) \right\rbrace+g_{12}
\left(\hat{\sigma}_{1}^{\dagger} \hat{\sigma}_{2}+ \hat{\sigma}_{1}
\hat{\sigma}_{2}^{\dagger}\right).
\end{equation}

\begin{figure*}[t]
\centering
\includegraphics[width=0.9\textwidth]{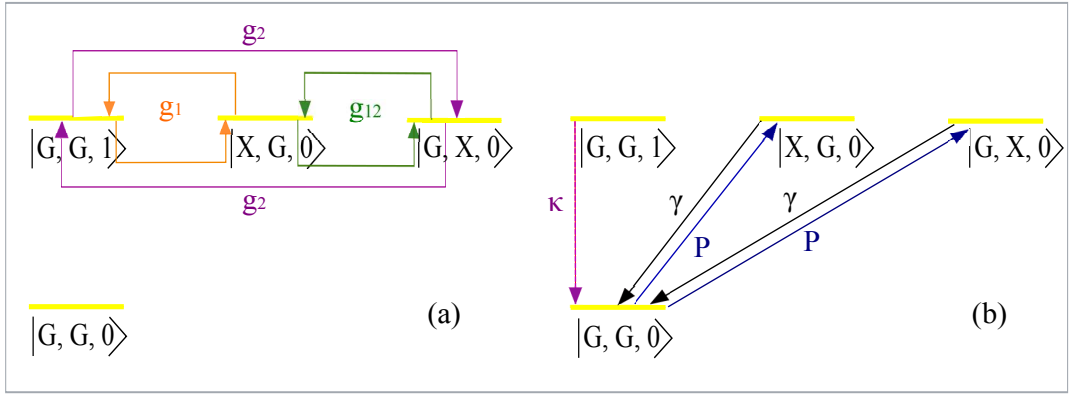} 
\caption{Dynamics of a system of two QDs in mutual interaction and interacting with a field mode, in presence of dissipative and pumping mechanisms. Considering up to one excitation. (a) The Hamiltonian dynamics $\hat{H}_S$. (b) Non-Hamiltonian dynamics.}
\label{fig0}
\end{figure*}

The first term corresponds to the free field Hamiltonian. The creation $\hat{a}^\dagger$
and annihilation $\hat{a}$ operators, are associated with photons of energy $\omega_0 = \omega_{X} - \Delta$ (where $\Delta$ is the detuning of each exciton respect to the field mode). The first term on the sum is related to the exciton energy. The operators
$\hat \sigma^{\dagger}_i=\ket{X_i}\bra{G_i}$ and $\hat \sigma_i=\ket{G_i}\bra{X_i}$ are
the creation and annihilation operators of the $i$th exciton. The exciton is
modeled as a two level system, where the ground state corresponds to the
absence of excitations and is denoted by $\ket{G}$, whereas the presence of
an excitation in the QD \textit{i.e} the exciton will be denoted as $\ket{X}$.
The transition energy between these states in any QD is $\omega_{X}$. 
The second term in the sum describes the dipolar interaction between the QDs
and the light mode \cite{Scully} in the rotating
wave approximation (RWA). The strength of such interaction is given by $g_i$. 
Finally, the last term accounts for the F\"orster exciton-exciton interaction, with coupling
constant $g_{12}$. This interaction represents the resonant exchange of
energy between the QDs.\\
On the other hand, the system-reservoir interaction Hamiltonian accounts for the
next processes: (i) The direct coupling between each exciton and the dispersive
photonic modes, this process is responsible for the spontaneous emission. (ii) The escape of the cavity mode photons, the so called coherent emission. (iii) A common continuous and incoherent pumping of each of the excitons.
The reservoir Hamiltonian $H_R$ contains the features of the
environment which can be modeled as a set of harmonic oscillators. It is not
necessary to write them explicitly. For a detailed discussion refer to the
literature of open quantum systems \cite{petruccione}. Finally to make the
problem tractable it is assumed that the interaction between the system and the
reservoir is weak, and that the dynamics of the reservoir is memoryless,
\textit{i.e} the Born-Markov approximation. Under these conditions, the
evolution of the reduced density operator of the system $\rho_S$ can be written
as:  
\begin{eqnarray}\label{eq.2}
\frac{d }{dt}  \hat{\rho}_{S} &=& i \left[\hat{\rho}_{S}, \hat{H}_{S} \right]
\nonumber \\
&+& \frac{\kappa}{2} \left(2  \hat{a} \hat{\rho}_S \hat{a}^{\dagger} - \hat{a}^{\dagger} \hat{a} \hat{\rho}_S -\hat{\rho}_S \hat{a}^{\dagger} \hat{a} \right) \nonumber \\
&+& \frac{\gamma}{2} \sum_{i=1}^{2} \left( 2 \hat{\sigma}_{i} \hat{\rho}_S \hat{\sigma}_{i}^{\dagger} - \hat{\sigma}_{i}^{\dagger} \hat{\sigma}_{i} \hat{\rho}_S - \hat{\rho}_S \hat{\sigma}_{i}^{\dagger} \hat{\sigma}_{i}\right) \\ 
&+& \frac{P}{2} \sum_{i=1}^{2}  \left( 2 \hat{\sigma}_{i}^{\dagger} \hat{\rho} \hat{\sigma}_{i} - \hat{\sigma}_{i} \hat{\sigma}_{i}^{\dagger} \hat{\rho} - \hat{\rho} \hat{\sigma}_{i} \hat{\sigma}_{i}^{\dagger}   \right) \nonumber ,
\end{eqnarray}
where $\kappa$ and $\gamma$ are the coherent and spontaneous emission rates respectively, and $P$ is related to the incoherent pumping rate of each exciton. The above parameters will be called dissipative parameters through the whole text.\\
The dynamics of the populations and coherences of the density matrix can be
obtained from equation (\ref{eq.2}).
For the results to be presented below we use the bare states basis: $\lbrace \vert k_1 \rangle
\otimes \vert k_2 \rangle \otimes \vert n \rangle \rbrace$, which we will denote
for convenience as: $\lbrace \vert k_1, k_2, n \rangle \rbrace$. In this basis
$\vert k_i \rangle$ represents the excited $\vert X \rangle$ or ground $\vert G
\rangle$ states of the \textit{i-th} exciton, whereas $\vert n \rangle$
indicates a Fock state with $n$ photons. In figure \ref{fig0} we show an schematic 
representation of the dynamics of the system. In figure 1(a) we portray all the possible 
transitions between the states of the system up to the first excitation manifold, 
associated to the Hamiltonian given by equation (\ref{eq.1}). It is seen that the 
Hamiltonian dynamics only admits horizontal transitions between states of the same 
excitation manifold, and there is no coupling between states belonging to different 
excitation manifolds. On the other hand, when we take into account the dissipative terms, 
see figure \ref{fig0} (b), the non Hamiltonian dynamics couples states 
between different excitation manifolds. Notice that, if we consider only the terms 
$\kappa$ and $\gamma$, the stationary solution will always be
the vacuum state $\vert G, G, 0 \rangle$. Nonetheless, if we introduce the common 
pumping term the system can have as stationary solution populations of excitons or 
photons different from zero. This is a crucial fact in the determination of the SC, 
as can be seen from the results of \cite{Elena08}. 

\section{First order correlation function and spectrum}
\label{S3}
As we mentioned in section \ref{S1}, the characterization of the dynamical
regimes can be obtained from the emission spectrum of the system. To obtain the spectral function
we use the Wiener-Khintchine theorem. The emission spectrum of the cavity can be obtained from the first order correlation function of the field: $S (\omega) \propto \lim_{t\to\infty} \Re \left[ \int_{0}^\infty d\tau e ^{i \omega \tau} G^{(1)} (t, \tau) \right] $, where $G^{(1)} (t, \tau) \doteq \langle \hat{a}^\dagger (t)
\hat{a} (t + \tau) \rangle $ is the first order correlation function.
The dynamics of the first order correlation function can be obtained through the quantum regression theorem 
(QRT) \cite{Scully}, which states that once the evolution of a set of operators $\lbrace \hat{O_i}
(t)\rbrace$ of the form $\partial_\tau \langle \hat{O}_j (t + \tau) \rangle =
\sum_j A_{ij} \lbrace \hat{O}_j (t + \tau) \rbrace$  is known, then, the two
time expected values of $\hat{O}_j$ with any operator $\hat{O}$ also satisfies
the same system of differential equations, \textit{i.e} $\partial_\tau \langle
\hat{O}_j (t + \tau) \hat{O} \rangle = \sum_j A_{ij} \lbrace \hat{O}_j (t + \tau
) \hat{O} (t) \rbrace$. We will use the QRT to obtain the equation of motion of
$G^{(1)}(t, \tau)$, and by studying the eigenvalues of the matrix that represents 
the two-time dynamics, which we call the QRT-matrix, we will introduce a SC
criterion. To do so, we consider the evolution given by equation (\ref{eq.2})
including states of up to one excitation. This set corresponds to the
$\lbrace \vert G, G, 0 \rangle , \vert G, G, 1 \rangle, \vert X, G, 0 \rangle\ ,
\vert G, X, 0 \rangle \rbrace$ states, which are depicted in figure \ref{fig0} . 
We select the set of operators: $\lbrace \hat{a}^{\dagger}, \hat{\sigma}_1^{\dagger}, \hat{\sigma}_2^{\dagger}
\rbrace$, and calculate the dynamics of this operators as 
$ \langle \hat{O} \rangle = \tr[\hat{O} \hat{\rho}_S]$. Finally, 
we find the QRT-matrix for the two time expected values.
\section{Results}
\label{S4}
\begin{figure*}[t]
\centering
\includegraphics[width=0.8\textwidth]{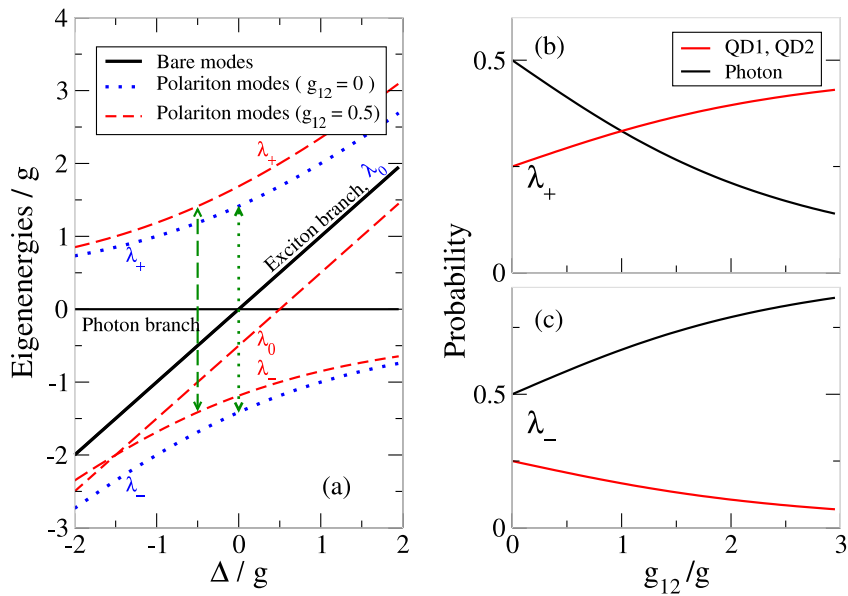} 
\caption{(a) Left. Eigenenergies of $H_S$ as a function of the detuning $\Delta$. The black continuous lines represent the photon mode energy (fixed at $\omega_0 = 0$ for simplicity) and the exciton energy $\omega/g = \lambda_0\vert_{g_{12}=0}$. In blue dotted lines the energies $\lambda_{\pm}$ are plotted for the case $g_{12} = 0$, as we said before they are above and below from the energy value $\lambda_0$. In red dashed lines we plotted the energies when the interaction parameter is turned on $g_{12} = 0$.$5 g$. Green arrows indicate the minimum distance approach between the $\lambda_{\pm}$ energies, dashed arrow for the case $g_{12} = 0$.$5 g$ and dotted arrow for the non interacting case $g_{12} = 0$. Right. Eigenvectors associated to its eigenvalues  $\lambda_{+}$ upper panel (b), $\lambda_{-}$ bottom panel (c). The red lines indicate the contributions of both of the QDs, and the black line shows the contribution of the light mode to the states of the coupled system, as a function of the interaction between the QDs.}
\label{fig1}
\end{figure*}

To understand the nature of the transitions and the emission peaks of the
system we first study it without considering the influence of
external reservoirs.  We obtain the polaritonic or dressed states and energies of
the Hamiltonian $H_S$, by consider $g_1 = g_2 = g$, (This ideal situation corresponds to the assumption that the excitons have equal coupling strengths to the light). The energies are: 
\begin{equation}\label{eq.3}
\lambda_0 =  \omega_0 + \Delta - g_{12}, \hspace*{0.3cm} 
\lambda_{\pm} = \left( \omega_0 + \frac{ g_{12} + \Delta}{2} \pm \mathcal{R} \right) ,
\end{equation}
\noindent
where $\mathcal{R} = \sqrt{2 g^2+ \left(( \Delta + g_{12})/2 \right)^2}$, this result has an important meaning. If the initial state has one matter excitation then the squared modulus of the transition probability to a state with a single photon is: 
\begin{eqnarray}
|\langle G, G, 1 \vert e^{- i H_S t} \vert X, G, 0 \rangle |^2 &=& |\langle G, G, 1 \vert e^{- i H_S t} \vert G, X, 0 \rangle |^2  \\ \nonumber
&=& \frac{g^2}{2 \mathcal{R}^2} (1 - \cos \left[ 2 \mathcal{R} t \right]  ),
\end{eqnarray} 
therefore the squared modulus of the transition probability oscillates with frequency $2 \mathcal{R}$, this can be understood as the \textit{Rabi frequency} of this system. In the case in which there is no mutual interaction between the excitons, $\mathcal{R}$ reduces to $\mathcal{R} = \sqrt{2 g^2 + \left( \frac{\Delta}{2} \right)^2 }$, which except for a factor of two, due to the fact that we are considering two excitons, is the usual \textit{half Rabi frequency} \cite{Gerry}. Also notice that in addition to the bare detuning factor $\Delta$, the term $g_{12}$ enters as an extra detuning factor.
In figure \ref{fig1}(a) we plot the energies as a function of the detuning $\Delta$. The bare mode energies are shown in black lines, the cavity mode energy remains constant, while the QDs exciton energies change with the detuning, and intersect with the cavity mode energy at resonance. 
The energies $\lambda_{0,\pm}$ which correspond to equation (\ref{eq.3}), are shown in colors for two different situations. First when there is no interaction between the QDs \textit{i.e} $g_{12} = 0$ (blue dotted lines),
second, when the interaction parameter is turned on, in this case we set
$g_{12}= 0.5 g$ (red dashed lines). In both cases the lines associated to $\lambda_{\pm}$ never cross each other. Nonetheless, when the QDs interaction is turned on, its effect is to slightly modify
all the energies, moving upwards both $\lambda_{\pm}$, and moving downwards $\lambda_{0}$, which causes a crossing with $\lambda_{-}$. Yet, another prominent aspect of the interacting case is that the minimum approach distance between the energies $\lambda_{\pm}$ occur off resonance (green dashed arrow), unlike the non interacting case (green dotted arrow). The separation in energy can be calculated as: $\Delta E = \lambda_+ - \lambda_- = 2 \mathcal{R}$, which is a minimum for $g_{12} = - \Delta$ . Therefore, the off-resonance minimum separation energy associated to the upper and lower polariton modes, can be seen as a witness of the interaction between the pair of QD excitons. Despite the demanding experimental conditions in the control of the interaction between the QDs, the value of this interaction constant can be at least in principle determined in a simple experimental way. 

On the other hand, the form of the eigenstates of $H_S$, is given by:
\begin{eqnarray}\label{eq.4}
 \ket{\lambda_0}&=&\frac{1}{\sqrt{2}}\left(\ket{G,X,0}-\ket{X,G,0}\right), \nonumber\\
 \ket{\lambda_\pm}&=&\frac{1}{N_{\pm}} \left(- \frac{1}{2g} \left\{ \Delta + g_{12} \pm 2 \mathcal{R} \right\} \ket{G,G,1}+\ket{X,G,0}+\ket{G,X,0} \right),
\end{eqnarray}
\noindent
where $N_{\pm}$ is a normalization factor. It is seen that $\ket{\lambda_0}$ 
is a maximally entangled exciton-exciton Bell state that is also completely separable 
from the light state. This is a result of choosing the symmetric condition $g_1 = g_2 = g$. 
On the other hand the states $\ket{\lambda_{\pm}}$ have more complex entanglement properties, 
for instance, in the case where $g=g_{12}$ the state $\ket{\lambda_-}$ reduces
to the 3 qubit W-state \cite{Wstate}. In figure \ref{fig1} (b), (c),  we plot the contributions of the QDs and the cavity mode to the quantum states of the coupled system (\textit{i.e} the polaritons) for each energy $\lambda_{\pm}$ at resonance as a function of the relative interaction strength $g_{12}/g$.
The case $\lambda_0$ was not plotted because the occupations correspond to the maximally entangled exciton-exciton Bell state, and they do not depend on $g_{12}$. The upper and bottom panels are related to the eigenenergies $\lambda_{\pm}$ respectively. The black line corresponds to the occupation of the light mode, whereas the red line corresponds to the QDs occupations, which is the same for both of them because of the symmetry conditions imposed. From the upper panel (b), we can see  that at $g_{12} = 0$, the system has photon-like and exciton-like components, at $g_{12} = g$ we find a complete mixture of all three states, for greater values of $g_{12}/g$ the system goes to an exciton-like state. In the bottom panel (c) we find the same condition as before for $g_{12} = 0$, in this case the system goes to have an strong photon-like component, and lesser exciton-like component as $g_{12}/g$ increases.
Now, we will obtain the emission spectrum for the system, this requires to take into account the dissipative effects. To do so, we derive the dynamical equations of the first order correlation functions $\textbf{v} (t +
\tau, t)= \{\langle \hat{a}^{\dagger}(t + \tau) \hat{a} (t)\rangle,\langle
\hat{\sigma}_{1}^{\dagger}(t + \tau) \hat{a} (t) \rangle,\langle
\hat{\sigma}_{2}^{\dagger}(t + \tau) \hat{a} (t) \rangle \} $, by using the QRT.
We obtain a linear system of the form : 
$\frac{d}{d\tau} \mathbf{v} (t +\tau,t)= \mathbf{A} \mathbf{v} (t +\tau,t)$
with:
\begin{eqnarray}{\label{eq.5}}
\mathbf{A}=
 \left(
\begin{array}{lll}
-2P -\frac{\kappa}{2} + i \omega_{0} & i g_{1} & i g_{2} \\
i g_{1} & -\frac{1}{2} (2P + \gamma) + i \omega_{X} & i g_{12} \\
i g_{2} & i g_{12} & -\frac{1}{2} (2P + \gamma) + i \omega_{X} \\
\end{array}
\right),
\end{eqnarray}
where the parameters $P,\kappa, \gamma$, \ldots, are those already introduced in equations (\ref{eq.1}, \ref{eq.2}). The dynamical evolution of $\textbf{v} (t + \tau, t)$ is given by: $\textbf{v}
(t + \tau, t) = e ^{\textbf{A} \tau} \textbf{v} (t,t) $. Even for the symmetric case the complete solution is rather cumbersome and is not presented here.
To obtain the criterion for the coupling regimes, we begin studying the eigenvalues $\alpha_{0,\pm}$ of the matrix $\mathbf{A}$, which are related to the positions $\Omega_{0,\pm}$ and widths $\Lambda_{0,\pm}$ of the spectrum ($i \alpha_{0,\pm} = \Omega_{0,\pm} + i \Lambda_{0,\pm} $), we find:
\begin{eqnarray}\label{eq.6}
 \alpha_0 &=& i \left( \omega_0 + \Delta - g_{12} \right) - \frac{\gamma}{2} - \frac{3 P}{2} , \\
 \alpha_\pm &=& i \left( \frac{g_{12} + \Delta}{2}  + \omega_0 \pm R \right)  - \frac{7P + \gamma + \kappa}{4} , \nonumber
\end{eqnarray}
where:
\begin{equation}\label{eq.7}
R  = \sqrt{2 g^{2} - \left( \Gamma + i \left( \frac{\Delta + g_{12}}{2}\right) 
\right)^2 }, \quad 
\mbox{with} \quad \Gamma = \frac{P + \kappa - \gamma}{4}.
\end{equation}

The number $R$ has a similarity with the \textit{Rabi frequency}, found at the beginning of this section, we called it \textit{half complex Rabi frequency}, and it includes the dissipative effects. Notice that if $R$ has a non-zero real part, the positions of the emission peaks $\Omega_{\pm}=\Re(i \alpha_{\pm})$ are different. On the other hand, if $R$ is a pure imaginary number, the contributions to $\alpha_{\pm}$ will merely affect the width of peaks of the spectrum. Based on the last statement, we introduce a SC criterion in presence of dissipation as follows:

\textit{The SC or WC regimes for a system of two excitons interacting with a confined light mode, is defined as the regime for which the complex Rabi frequency is a purely real (SC) or purely imaginary (WC) number, under the off-resonance condition $\Delta = - g_{12}$}.

\begin{figure*}[t]
\begin{center}
\begin{minipage}{8cm}{
\centering
\begin{center}
$|\Re(R)/\Im(R)|$
\end{center} 
\includegraphics[scale=1]{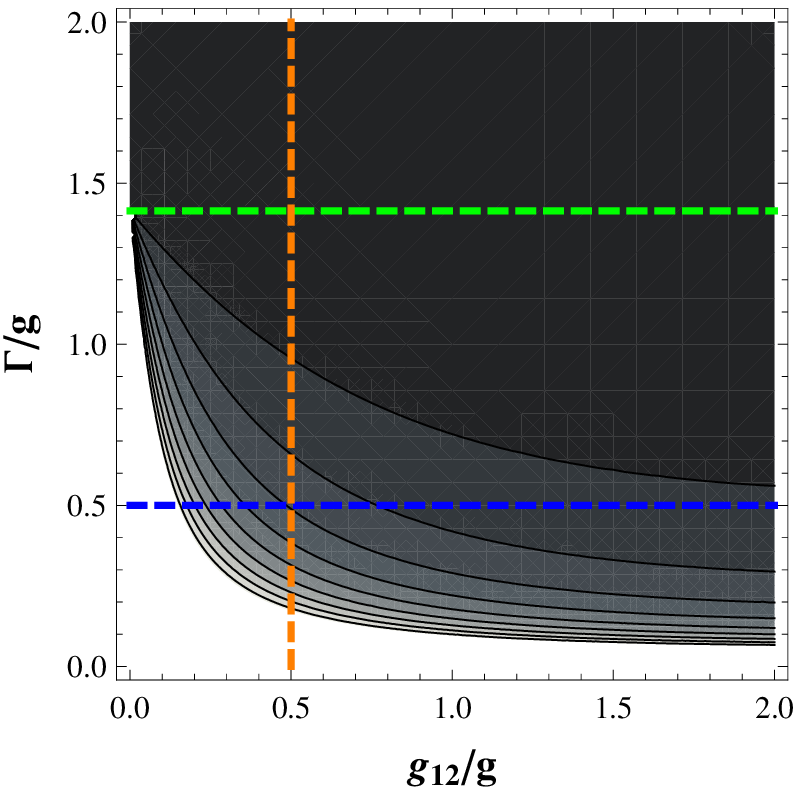} 
}
\end{minipage}
\begin{minipage}{4.5cm}{\begin{center} 
\includegraphics[scale=1]{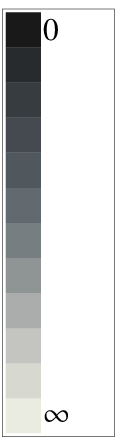} 
\end{center}}
\end{minipage}
\end{center} 
\caption{Contour plot for the ratio $\vert \Re(R)/\Im(R) \vert$ between the
real and imaginary parts of the complex Rabi frequency $R$ at resonance, as a function of the
decay rates $\Gamma$ and the coupling constant $g_{12}$ between
the QDs. The SC regime is obtained for parameter values such that $\vert
\Re(R)/\Im(R) \vert \rightarrow \infty$. The WC regime is reached when $\vert
\Re(R)/\Im(R) \vert \rightarrow 0$. Green dashed line represent the upper limit for the SC regime. Orange, and blue
dashed lines are used to represent some parameters for which we plot the emission spectrum in figure \ref{fig3}.}  	 
\label{fig2}
\end{figure*}

\begin{figure*}[t]
\centering
\includegraphics[width=0.8\textwidth]{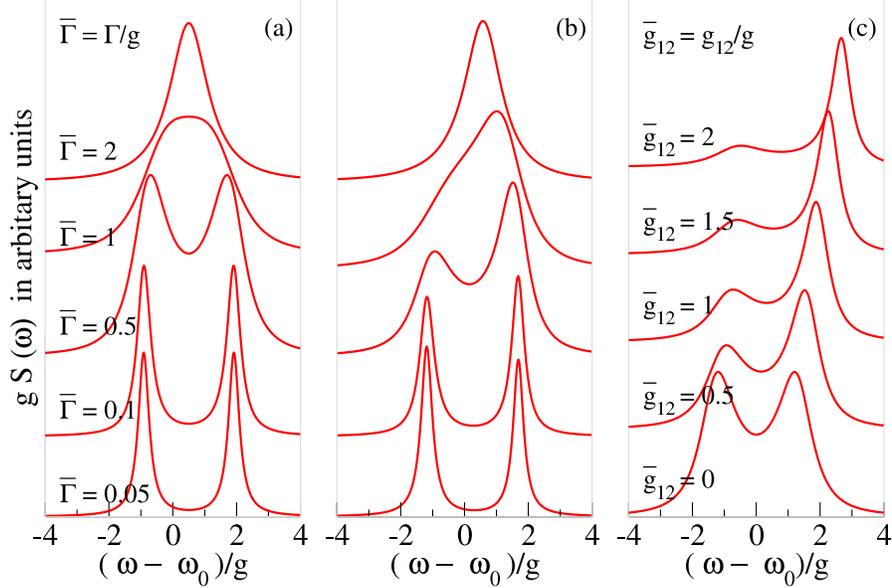} 
\caption{Emission spectrum at resonance as a function of $\omega - \omega_{0}$ in arbitrary units. (a) Emission
spectrum for both: the resonant condition and without interaction between the dots $g_{12} = 0$, and the \textit{re-normalized complex Rabi frequency}. (b) Spectrum for different dissipative values, keeping fixed $g_{12} =0.5g$, which is related to change  $\Gamma$ over the orange line in figure \ref{fig2}. (c) Spectrum for different values of $g_{12}/g$, in this case the dissipative parameters were keep constant $\Gamma = 0.5g$, and corresponds to the blue line in figure \ref{fig2}.} 
\label{fig3}
\end{figure*}

Consequently, the system reaches the SC regime if $\Gamma/g < \sqrt{2}$, this condition is represented by the region below the green dashed line in figure 3. However, to determine the coupling regime, usually the experimental spectra is analyzed near resonance. In this case the behaviour of the real ($\Re(R)$) or imaginary part ($\Im(R)$) of the complex Rabi frequency $R$, screens the dynamical regime of the system. This can be seen in figure \ref{fig2}, in which we present a contour plot of the ratio $|\Re(R)/\Im(R)|$ at resonance, as a function of the two free parameters: $\Gamma/g$ and $g_{12}/g$.  The predominance of $\Im(R)$ is represented in dark colors ($|\Re(R)/\Im(R)|\rightarrow 0$, a WC-like character), on the contrary the bright colors are associated to the predominance of $\Re(R)$ ($|\Re(R)/\Im(R)| \rightarrow \infty$, a SC-like character). Note that, even below the green dashed line, $\Im(R)$ prevails and therefore in these conditions the characteristic two peaks feature of the SC regime is lost.
Finally, to check that the criterion we have defined agrees among the
different coupling regimes, in figure \ref{fig3} we plot the normalized emission spectrum for three different cases. In plots (a), (b) we choose variations in $\Gamma/g$, the dissipative parameters, whereas in (c) we take into account variations in $g_{12}/g$, represented along the vertical axis. For the case $\Delta = g_{12} = 0$ the emission spectra is presented in (a). The SC features arise for small values of the dissipative parameters and the splitting between the peaks is clearly observed. The symmetry in the peaks distribution is caused by the selected parameters, where both excitons are coupled equally to the cavity. On the other hand, as the dissipation parameter $\Gamma/g$ increases, the peaks become broader and closer. Finally, for high values of $\Gamma/g$, the SC behavior is lost, then the system is in the WC regime.

It is clear from equation (\ref{eq.7}), that the effect of $\Delta$ and $g_{12}$ is to increase the splitting between the peaks, and therefore the SC condition can be reached by controlling these two parameters. However in figure \ref{fig3} (b), we plot the emission spectra at resonance and $g_{12} = 0.5g$ (represented by the vertical orange line plotted over figure \ref{fig2}). As before, we found the SC regime for small values of the dissipative parameters, and eventually the WC is reached for sufficiently high values of dissipation. 
Notice in figure \ref{fig3} (a), (b) that for the dissipative parameters $\Gamma = 1$, the two peaks feature is lost, but the system is in the SC regime (from the experimental point of view, this kind of broad spectra is deconvoluted into two components). Thus, again we can see that at resonance the effect of the interaction is to screen the SC regime of the system.

Finally in figure \ref{fig3} (c) we keep constant the dissipative parameters (represented by the blue
line over figure \ref{fig2}). When $g_{12} = 0$, the spectrum is completely symmetric. As $g_{12}$ increases, the splitting between the peaks increases as well, and one of the peaks decreases in intensity. This change in the intensity, can be explained by inspecting the behavior of the polaritonic states $\vert \lambda_{\pm} \rangle$ as a function of $g_{12}$. From figure \ref{fig1} (b) we recall that $\vert \lambda_{-} \rangle$ acquires a strong photonic-like character as $g_{12}$ increases. On the other hand $\vert \lambda_{+} \rangle$ becomes more exciton-like; therefore the state $\vert \lambda_{-} \rangle$ decouples progressively from the transitions induced by the matter-matter interaction.  

\section{Discussion and Conclusions}
\label{S5}

From the above results, it is interesting to notice that the off-resonance minimum separation between the polaritonic energies, can be used as a witness of the mutual interaction between the QDs, \textit{i.e} the F\"orster interaction, and in fact this coupling interaction constant can be measured directly. On the other hand, the most relevant result of this work is that we have established a criterion for the different dynamical regimes for a system of two interacting excitons symmetrically coupled to a photonic mode in a semiconductor microcavity. This criterion
is based on the real and imaginary parts of the complex Rabi frequency $R$. In
accordance to experimental results of SC for this systems, it has been
observed that the SC regime is reached for small values of $\Gamma$ relative to
$g$, the light matter coupling constant. This system will be in the SC regime
for values of $\Gamma /g \leq \sqrt{2} $ (see the green dashed line in \ref{fig2}). 
On the contrary for values of $\Gamma /g \geq \sqrt{2} $ we get WC independently of the interaction between the excitons. The above results could indicate that the experimental conditions to achieve
Bose-Einstein condensation of polaritons (SC) or single photon sources (WC) are
not strongly influenced by the exciton interaction. 
Finally from the expression of the complex Rabi frequency equation (\ref{eq.4}),
it is shown that the effect of the number of particles in the multiexcitonic
system has two opposite effects. First, the mutual excitonic interaction is not
favorable to the SC because it takes $R$ to the limit of a purely imaginary
number. In second place, the collective effects intensify the SC as the term
proportional to $g$ increases, in fact this term scales as a typical Tavis-Cummings factor $\sqrt{N} g$ \cite{Tavis} where $N$ is the number of excitons assuming the dots are identical, as we expected for a lineal regime. \\

\ack
We gratefully thank Professor P.S. Soares Gimar\~aes and Professor K. Fonseca for their useful discussions. This work was supported by CODI at Universidad de Antioquia and Universidad Nacional de Colombia.

\section*{References}

\bibliographystyle{iopart-num}
\bibliography{SC}

\end{document}